\documentclass[11pt,twoside]{article}
\usepackage{asp2004}
\usepackage{psfig}
\usepackage{epsf}
\usepackage{graphics}
\usepackage{lscape}
\def\edcomment#1{\iffalse\marginpar{\raggedright\sl#1\/}\else\relax\fi}
\reversemarginpar

\begin{document}
\title{A Project for a New Echo-Mapping Campaign\\ at Intermediate Redshift}
\author{Dario Trevese$^1$, Giovanna Stirpe$^2$, Fausto Vagnetti$^3$, 
Valentina Zitelli$^2$, and Diego Paris$^1$}
\affil{$^1$Universit\`a di Roma La Sapienza, Dipartimento di Fisica, p.le Aldo Moro 2, I-00185 Roma, Italy}
\affil{$^2$INAF - Osservatorio Astronomico di Bologna, Italy}
\affil{$^3$Universit\`a di Roma Tor Vergata, Dipartimento di Fisica, via della Ricerca Scientifica 1, I-00133 Roma, Italy}

\begin{abstract}
AGN masses  can be estimated  by "single epoch"  spectral measurements
through  a  mass-luminosity-
linewidth  relation calibrated  by  echo
mapping  measurements of  a reference  sample
of low  redshift  ($z <
0.3$) and low luminosity ($M_B  > -26$) AGNs.
To analyze the possible
dependence  of  this  relation  on  luminosity we  selected  a  sample
of
bright, intermediate redhift ($z \sim 1$) objects and we started a
spectrophotometric monitoring
campaign  with a typical  sampling time
of about  one month.  Variability observations of  lines with
shorter
wavelength  than H$_\beta$ will  also provide  new information  on the
structure  of the  broad
line region.  Cross-correlation  analysis of
continuum and  line variations  will require years  of
monitoring. We
present a preliminary analysis of  the data collected during the first
year   of
observations   and  we   discuss   the   adequacy  of   the
spectrophotometric  accuracy attained  and  future
prospects of  this
project.
\end{abstract}
\thispagestyle{plain}

\section*{Introduction}
Broad  emission lines  in  AGN  spectra respond  to  variation of  the
central source  continuum luminosity with a delay  due to light-travel
time effects.   This allows to derive  the geometry of  the broad line
region   (BLR)  through   the   "echo  mapping"   process,  based   on
cross-correlation analysis  of continuum and  line variations. Studies
of  Seyfert galaxies  provided sizes $R_{BLR}$ of  the BLR  from light-days  to
light-weeks, an  order of magnitude smaller  than originally predicted
by photoionization  models, providing  new insight
into  the  physics  of  BLR  gas (Peterson  1993;  Netzer \& Peterson
1997).
Adding to the geometrical information an estimate
of cloud velocity, derived from the width of $H_{\beta}$ emission line
$\Delta{H_{\beta}}$, it is possible to
derive  the mass  $M_{BH}$ of the central black hole as
$ M_{BH} = {c_o} \cdot R_{BLR} \cdot \Delta H_{\beta}^2$.
The analysis of a sample of 34 AGNs with $B < 16$ and
$z < 0.4$, observed  partly  at  the   2.3  m  telescope  of  the  Steward
Observatory and partly  at the 1 m telescope  of the Wise Observatory,
allowed the extension  of the sample up to  100 times brighter objects
and the definition of a size-luminosity relation
of the type $R_{BLR}=c_1 L_{\lambda}^{\gamma}$, resulting in a mass-luminosity
relation:
$M_{BH} = {c_2} \cdot L_{\lambda}^{\gamma} \cdot \Delta H_{\beta}^2$
(Kaspi et al. 2000).
The latter relation can be used to derive "single epoch" mass estimates
based on measurements of $L_{\lambda}$ and $\Delta H_{\beta}$, allowing,
in principle, to study the evolution of AGN masses in cosmic time.
However this  requires an
extrapolation of the above relation to high-redshift and high-luminosities
objects.
The reliability of the extrapolation depends on the possibility of a
calibration with masses and luminosities directly derived from
echo-mapping measurements of high-$z$ and high-$L$ objects, which are observed
at shorter rest-frame waveleghts where different emission lines are seen
(e.g C IV 1540 \AA\ instead of Hydrogen Balmer lines).

For this purpose we have started a spectrophotometric
monitoring campaign of a sample of 
bright, intermediate redshift objects, using  the 1.5 m and the 1.8 m
telescopes, respectively  of the
Loiano (Bologna) and  Asiago (Padua) Observatories.

\section*{Sample and Observations}

Selection from the Veron-Cetty \& Veron (2003) catalog of all AGNs brighter
than $V=15.7$, $z >1$ and $\delta_{2000}>0$  yields 12 objects with
redshifts  in the range
1.083 to 3.911. We started  repeated observations of 5 of them, listed
in  Table 1 together  with their  positions, redshifts,  apparent $V$
magnitudes.  Both  Loiano and Asiago  telescopes are equipped  with
identical  Faint  Object Spectrograph  \&  Camera,  BFOSC and  AFOSC
respectively, designed to allow a quick switching between spectroscopic
and imaging mode.   BFOSC uses a back illuminated  1300x1340 CCD array
EEV D129915 with 20x20 micron  pixels corresponding to a scale of 0.58
arcsec pixel$^{-1}$.   We adopted  a 5"-wide slit  and a grism  with a
dispersion  of  3.97  \AA\  pixel$^{-1}$ corresponding  to  a  spectral
resolution of $\sim 10$ \AA\ and a spectral range 3800-8500 \AA .  

AFOSC  uses a  1100x1100 CCD
array TEK1024 Thinned with  22x22 micron pixels corresponding to 0.473
arcsec.  We adopted a  8".4-wide slit  and a  grism with  a 
dispersion  of 4.99  \AA\ pixel$^{-1}$, corresponding to  a resolution
$\sim 13$ \AA\ and a spectral  range 3500-8450 \AA .  
For  each  QSO  we selected  a
comparison star of  comparable magnitude included in a  field of about
15x15 arcmin (see Maoz et al. 1994), and spectroscopic  observations
were performed rotating the
slit to  include the comparison  star.
This  allows  the  normalize  the  data at  one  reference  epoch,  as
described  below.   Typical observations  consist  of two  consecutive
exposures of 1800 s. The QSO to reference star ratio, as a function of
wavelength  is  computed  for   each  frame.  This  quantity  must  be
independent  of extinction  changes during  the night.  This  allows a
check of consistency between the  frames and the rejection of the data
if inconsistencies occur.

Direct B  and R images in 15x15  arcmin field centered on  the QSO are
taken  to check  possible variability  of the  reference star  and the
consistency of spectrophotometry with broad band photometry.

\begin{center}
{\bf TABLE 1}\\
\begin{tabular}[t]{|lll|}
\hline
quasar        &$z$   &V\\
\hline
APM08279+5255 &3.911 &15.20\\
PG1247+268    &2.042 &15.60\\
PG1634+706    &1.337 &15.27\\
PG1718+481    &1.083 &14.60\\
HS2154+2228   &1.290 &15.30\\
\hline
\end{tabular}
\end{center}

\section*{Data Reduction}
The  spectra of each QSO and  the relevant reference star
were extracted with the standard IRAF procedures
 and calibrated in wavelength using  exposures of He-Ar
and Hg-Cd arc  spectra, with BFOSC and AFOSC  respectively, taken close
to each  spectroscopic exposure.  Data of two  consecutive exposure of
the reference  star and  the QSO  were added to  reduce noise.  Then a
continuum   of   the   star    was   derived  at each epoch. 
Defining $\mu_i(\lambda)  \equiv\frac{C_r(\lambda)}{C_i(\lambda)}$
where $C_r(\lambda)$ and $C_i(\lambda)$ are star continua
at a  reference epoch  and at  the generic  i-th  epoch 
respectively, we reduce  each QSO
spectrum   $S_i(\lambda)$  to   the  reference   epoch   by  computing
$S'_i(\lambda)=\mu_i(\lambda) \cdot S_i(\lambda)$.
An absolute flux calibration, which is not necessary for relative
variability measurements, was also applied at the reference epoch, 
based on the observation of a standard star. Since all spectra
are reduced to the reference epoch, the same conversion to physical units
was applied to all of them. Thus the procedure does not add further 
uncertainties to relative spectrophotometry.
However uncertainty of the absolute calibration itself is $\sim$ 10\%, i.e.  
less accurate than relative photometry discussed below.
As an example, the spectrum of PG 1247+268 at two epochs, 
normalized to the reference epoch, is shown in Figure 1. 

\section*{Results and Discussion}
Line and continuum spectra were measured at each epoch. For each line to be
monitored we selected line-free regions on both sides of the line
and defined  continuum values in 100 \AA\ bands. The line flux is defined 
as the integral of the flux above a straight line connecting the two continuum
values. Lines and continuum regions depend on the QSO redshifts.
Here we present preliminary result of the analysis of two objects
 PG 1247+268 and PG 1634+706.
In the case of the first  ($z=2.042$) we analyzed C IV $\lambda$1549 and
 CIII] $\lambda$1909 while in the case of the second ($z=1.337$) 
we analyzed CIII] $\lambda$1909 and MgII $\lambda$2798.
The relevant bands for continuum measurements,
selected on the basis of an inspection of the average spectra,
are reported in Table 2.
Figure 2 represents line and continuum light curves of PG 1247+268.
The number of observations is still too small to derive any reliable conclusion about variability.
However, we can evaluate the adequacy of observation and data reduction adopted for the program.
The relative rms variations $\sigma/F$ for the lines and relevant continua of both QSOs are reported in Table 2.
These variations include the intrinsic QSO variability during one year, thus they represent an upper limit to the uncertainty of our photometry.
Since typical continuum QSO variability is of the order of a few tenths of a magnitude in a timescale of $\la 1$ yr in the rest-frame, we will be able to measure intrinsic variations after one more year of observation.

\begin{center}
{\bf TABLE 2}
{\footnotesize
\begin{tabular}[t]{|ll|ccc|ccc|ccc|}
\hline
            &               & &CIV& & &CIII]& & &MgII& \\
            &               & cont&line&cont& cont&line&cont& cont&line&cont\\
\hline
PG 1247+268 &$\lambda_{obs}$&4500&4712&4850&5510&5807&6020&&&\\
            &$\sigma/F$     &0.05&0.03&0.06&0.04&0.03&0.05&&&\\
PG 1634+706 &$\lambda_{obs}$&&&&4232&4461&4623&6280&6538&6750\\
            &$\sigma/F$     &&&&0.01&0.06&0.02&0.02&0.06&0.02\\
\hline
\end{tabular}
}
\end{center}

\begin{figure}[!ht]
\plotfiddle{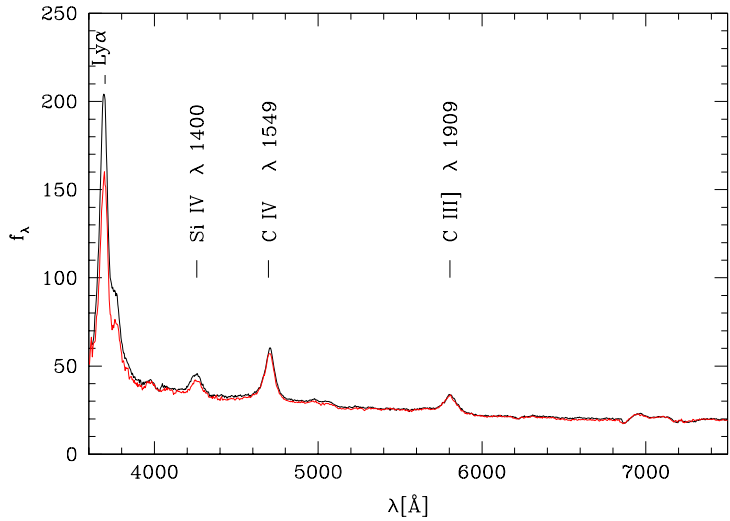}{6cm}{0}{50}{50}{-150}{-160}
\caption{Superposed spectra of PG 1247+268 observed  at two epochs,
at the two telescopes of Asiago and Loiano. $f_{\lambda}$ is in units of
10$^{-16}$ erg s$^{-1}$ cm$^{-2}$ \AA$^{-1}$.}
\end{figure}

\begin{figure}[!ht]
\plotfiddle{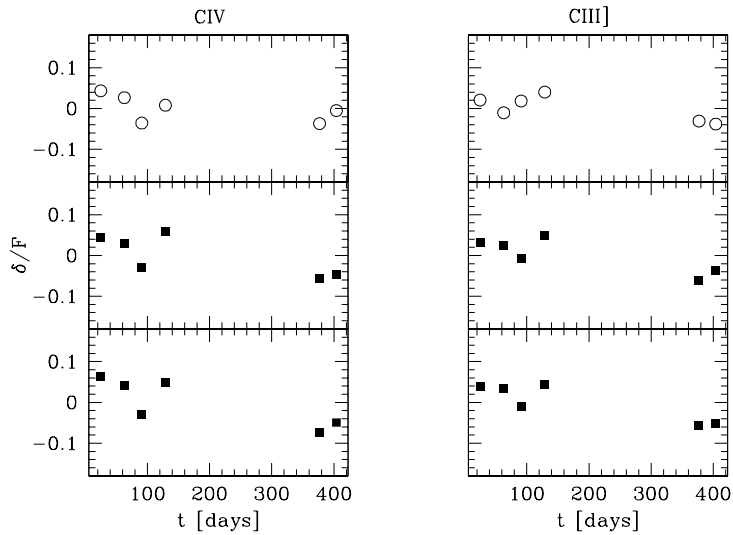}{6cm}{0}{50}{50}{-150}{-160}
\caption{Relative flux variations for PG 1247+268 emission lines
CIV $\lambda$ 1549 \AA\ and CIII] $\lambda$ 1909 \AA ~(open symbols)
and the relevant continua (filled symbols) as a function of time.}
\end{figure}

\end{document}